\documentclass[pre,twocolumn,superscriptaddress,amsmath,amssymb,showpacs,noshowkeys,a4paper]{revtex4-1}

\usepackage{amsmath}

\usepackage{graphicx}
\usepackage{dcolumn}
\usepackage{bm}
\usepackage[usenames]{color}
\usepackage[dvipsnames]{xcolor}
\usepackage{epstopdf}
\usepackage[normalem]{ulem}
\definecolor{mygray}{gray}{0.5}
\definecolor{amber}{rgb}{1.0, 0.49, 0.0}
\definecolor{ao}{rgb}{0.0, 0.5, 0.0}
\definecolor{azure}{rgb}{0.0, 0.5, 1.0}

\newcommand{\affGRASP}{\address{GRASP, UR CESAM, Institute of Physics B5a, Universit\'e de Li\`ege, B4000 Li\`ege, Belgium, EU}}
\newcommand{\affLeiden}{\address{Huygens-Kamerlingh Onnes Laboratory, Universiteit Leiden, PO box 9504, 2300 RA Leiden, The Netherlands, EU}}
\newcommand{\affJFI}{\address{James Franck institute, Department of Physics, University of Chicago, Chicago, IL 60637, USA}}

\begin{document}
\title{Self-propulsion and crossing statistics under random initial conditions}
\author{M. Hubert}
\affGRASP
\author{M. Labousse}
\affLeiden
\author{S. Perrard}
\affJFI
\email{sperrard@uchicago.edu}

\begin{abstract}
	We investigate the crossing of an energy barrier by a self-propelled particle described by a Rayleigh friction term. We reveal the existence of a sharp transition in the external force field whereby the amplitude dramatically increases. This corresponds to a saddle point transition in the velocity flow phase space, as would be expected for any type of repulsive force field. We use this approach to rationalize the results obtained by Eddi \emph{et al.} [\emph{Phys. Rev. Lett.} \textbf{102}, 240401 (2009)] who studied the interaction between a drop propelled by its accompanying wave field and a submarine obstacle. This wave particle entity can overcome potential barrier, suggesting the existence of a ``macroscopic tunneling effect". We show that the effect of self-propulsion is sufficiently strong to generate crossing of the high energy barrier. By assuming a random distribution of initial angles, we define a probability distribution to cross the potential barrier that matches with the data of Eddi \emph{et al.}. This probability is similar to the one encountered in statistical physics for Hamiltonian systems \textit{i.e.} a Boltzmann exponential law.
\end{abstract}
\pacs{47 55.D- Drops,  05 45.-a, Non linear dynamics and chaos}
\maketitle

\section{Introduction}

	Classical Hamiltonian systems are stuck at a given energy level and therefore cannot overcome barriers of potential energy. When one considers energy exchange with a thermal reservoir this property usually breaks down, as can be observed with Brownian motion and thermally activated processes. Self-propelled particles break the Hamiltonian structure and therefore may also overcome large potential barriers. This is a fundamental issue in active matter~\cite{Marchetti2013}, collective behaviors~\cite{Solon2015,Deseigne2010}, or motile colloidal systems~\cite{Bricard2013}.\\

	Here we investigate the possibility of self-propelled particles relying on Rayleigh friction~\cite{Rayleigh1877,Erdmann2005} to cross potential barriers. This nonlinear friction term was first introduced by Lord Rayleigh and has been since used for various motile systems~\cite{Bechinger2016,Romanczuk_EPJ_2012,Erdmann_2003,Kearns_2010}. The motility derives from an internal energy consumption input and an exchange with the environment, so that these particles may interact in a counter-intuitive way with external potentials. This model has been investigated in the case of thermally activated Brownian motion~\cite{Lindner2008} or in the presence of a quadratic~\cite{Erdmann2000}, cubic~\cite{Burada2012} and ratchet potential~\cite{Schweitzer2000}. \\
	
	 The present study is motivated by the experiments from A. Eddi \emph{et al.} \cite{Eddi2009} in which a walking droplet and its waves interact with a submarine obstacle, leading to ``classical tunneling'' of a wave-particle entity. Nachbin {\it et al.}~\cite{Nachbin2016} have investigated the case of one-dimensional crossings, where they have used a conformal mapping to model the two-dimensional flow below and to account for the presence of an immersed barrier. L. Faria~\cite{Faria2017} introduced an effective depth model which was subsequenlty used by Pucci {\it et al.}~\cite{Pucci2016} to model the non-specular reflection of a {\it walker}. \\
	 
	 In this article we adopt a complementary point of view and investigate whether the walkers crossing properties may be intrisic to their non-Hamiltonian nature.  In the short wave damping regime, the wave-drop association can be considered as a self-propelled particle experiencing a Rayleigh friction that originates from the waves emitted by the drop~\cite{Labousse2014,Bush2014}. We investigate whether a non-Hamiltonian particle can cross an energy barrier depending on its initial conditions. We  consider both linear and harmonic energy landscapes. In Sec.~\ref{SecII}, we present the theoretical model. Then in Sec.~\ref{SecIII}, we investigate the structure of the solutions. Using a representation in the velocity phase space, we show that a saddle point transition arises as a function of the external force amplitude. This transition separates two regimes that are qualitatively different in terms of crossing statistics. In the low force regime, we analytically derive the relationship between the incident angle and the maximal penetration depth of the particle. In Sec.~\ref{SecIV}, guided by the randomness of the impact angles observed during the crossing events in the experiment of Eddi \textit{et al.}, we add a stochastic feature by randomly choosing the initial conditions. We compare the probability, $\mathcal{P}$, to cross the barrier for random incident angles to the probability, $\mathcal{P}_{\mathrm{Boltz.}}$, that one would obtain from a thermally activated process, \textit{i.e.} Boltzmann exponential. We find $\mathcal{P}_{\mathrm{Boltz.}}$ = $\mathcal{P}$, which defines an equivalent temperature for the system. In Sec.~\ref{SecV}, we show that our model adequately captures the experimental observations of Eddi \emph{et al.}~\cite{Eddi2009}. Finally in Sec.~\ref{SecVI} we conclude and discuss the perspectives opened by this work. 
	
\begin{figure*}[t]
	\begin{centering}
		\includegraphics[width=0.74\textwidth]{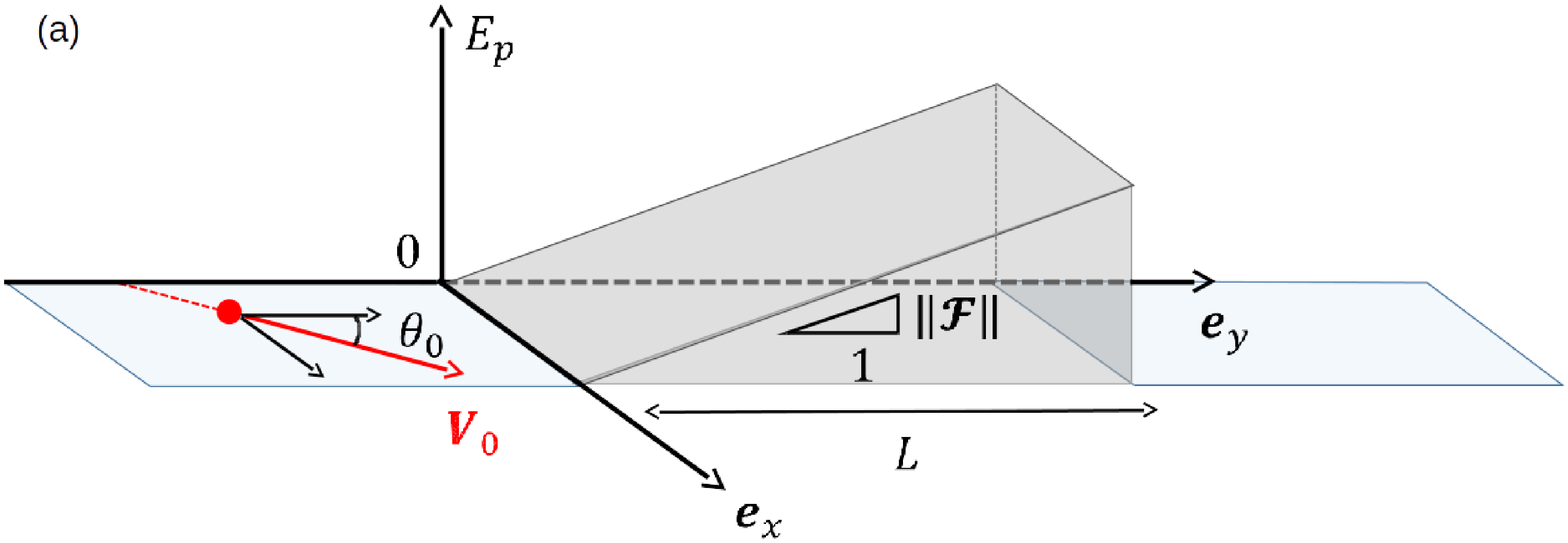}
		\includegraphics[width=0.25\textwidth]{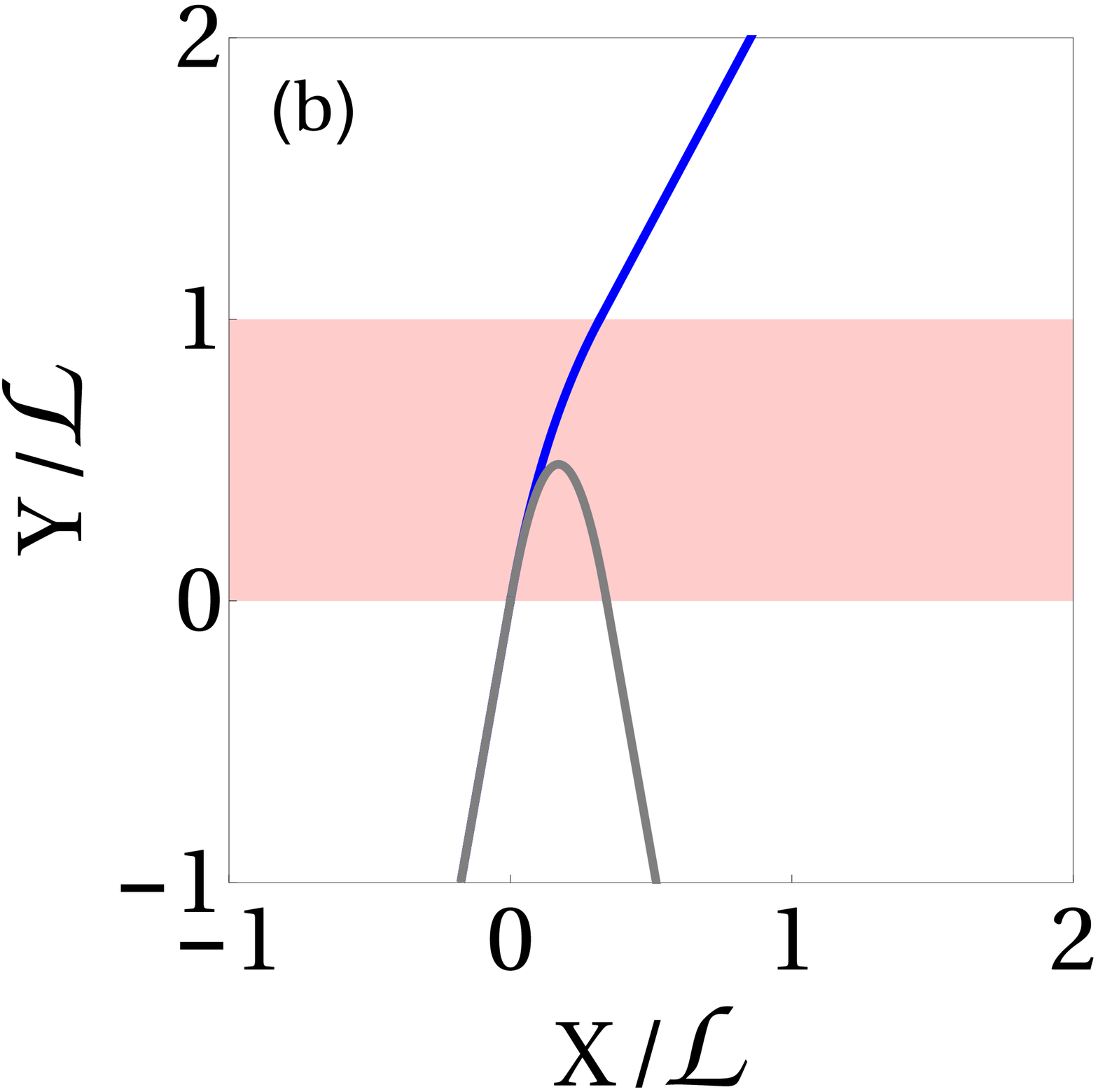}
		\caption{\emph{Color online}(a) Sketch of the thought experiment: The self-propelled particle, indicated by the red dot, moves with an initial velocity $\mathbf{v}_0 = (v_{x,0},v_{y,0})$ with $v_{y,0}$ pointing toward the potential barrier $E_p$ corresponding to a force field $\mathcal{F}$. The force field is applied in the interval $y\in[0,L]$ only. (b) Trajectories for two different constant force fields for $\theta_0 = 10^{\circ}$. In blue, $f = 10^{-3}$ and in gray $f = 10^3$. The red area indicates where the force field is applied. }
		\label{Fig1}
	\end{centering}
\end{figure*}

\section{Model\label{SecII}}
The model used throughout this article is the following. We consider a self-propelled particle of mass $m$ immersed in a two-dimensional force field, $\bm{\mathcal{F}}=- \mathcal{F} h_{[0,L]} \mathbf{e}_y$, that is constant in the region $Y \in [0,L]$ and zero elsewhere, as depicted in Fig.~\ref{Fig1}(a). Here, $h_{[0,L]}$ denotes a stepwise function of Y between $0$ and $L$ . This force is invariant along the x-axis. For the sake of simplicity, only the constant force field is fully investigated in this article, though we also study a harmonic potential $\bm{\mathcal{F}} = -m\Omega^2Y\mathbf{e}_y$, with $\Omega$ the angular frequency, which we will present shortly.\\
 
The self-propulsion is implemented by means of a Rayleigh-type friction force~\cite{Labousse2014} $\mathbf{F}_p$ reading
\begin{equation}
	\mathbf{F}_p=\frac{m}{T_v}\mathbf{V}\left(1-\frac{\Vert \mathbf{V}\Vert^2}{V_0^2} \right).
\end{equation}
Here, $\mathbf{V}$ the instantaneous velocity and $T_v$ the relaxation time toward the equilibrium velocity $V_0$. This term accounts both for an active propulsion and an effective friction that sets $V_0$. The force is propulsive if $V < V_0$ and the force leads to friction if $V > V_0$. This form was first introduced by Rayleigh~\cite{Rayleigh1877} and has been since applied to a wide range of systems such as self-propelled stochastic particles~\cite{Schweitzer2000}, car traffic~\cite{Helbing2001} or bouncing drops~\cite{Labousse2014}. Taking into account the force field $\bm{\mathcal{F}}$ and the self-propulsion $\mathbf{F}_p$, Newton's law for the self-propelled particle reads:
\begin{equation}
	\dot{\mathbf{V}}=\frac{1}{T_v}\mathbf{V}\left(1-\frac{\Vert \mathbf{V}\Vert^2}{V_0^2} \right)+\frac{\bm{\mathcal{F}}}{m}. 
	\label{equationofmotion}
\end{equation}
We use the dimensionless quantities $\mathcal{F} \rightarrow f= \mathcal{F} T_v/mV_0$, $t\rightarrow t/T_v$, $V \rightarrow v=V/V_0$ and consequently the spatial coordinates scales as $(X,Y)\rightarrow (x,y)=(X,Y)/(V_0 T_v)$. In the particular case of the harmonic potential we define a dimensionless angular frequency $\omega=\Omega T_v$.
The dimensionless equations of motion along $\mathbf{e}_x$ and $\mathbf{e}_y$ read
\begin{equation}
\left\{
    \begin{array}{l} 
    \dot{v}_x=v_x\left(1-\left(v_x^2+v_y^2\right) \right)\\
    \dot{v}_y=v_y\left(1-\left(v_x^2+v_y^2\right) \right)-f h_{[0,L]}
     \end{array}
\right.
\label{Equation2D}
\end{equation}
We solve this set of equations with Mathematica using the ``NDSolve'' algorithm. To compare regimes of very different force amplitudes, we introduce a force lengthscale $\mathcal{L}=mV_0^2/\mathcal{F}$ and express the results in terms of dimensionless distances $(X,Y)/\mathcal{L}$ or equivalently $(xf,yf)$. The corresponding lengthscale in the harmonic potential is given by{$\mathcal{L} = V_0/\Omega$. We define a penetration depth $L_c$ whose dimensionless form writes $\ell_c=L_c/\mathcal{L}$. Having introduced the model, we now investigate the possibility of crossing the potential barrier. \\

\section{Results: analysis of the two regimes $f < f^\star$ and $f > f^\star$  \label{SecIII}}
Fig.~\ref{Fig1}(b) illustrates the two regimes of propulsion that we identify. The particle trajectory is shown for two asymptotic force field magnitudes, $f =  10^3$ (grey line) and $f=10^{-3}$(blue line). The red shaded area corresponds to the region of space where the force field is applied. For the specific incident angle $\theta_0 = 10^{\circ}$ the particle crosses the constant force field provided $f>0.471$. This illustrates the existence of a transition from non-crossing to barrier crossing when the force field, $f$, is decreased. For small values of $f$, the particle only slightly deviates while maintaining its speed along the trajectory. \\
	
The transition between the two regimes of propulsion can be conveniently studied by considering the velocity potential:
\begin{equation}
	\mathcal{V}(v_x,v_y) = \frac{\Vert \mathbf{v}\Vert^2}{2}\left(\frac{\Vert \mathbf{v}\Vert^2}{2}-1\right)+fv_y.
\end{equation}
	The total dimensionless forces, $-\bm{\nabla}_{\mathbf{v}} \mathcal{V}(v_x,v_y)$, are the derivatives of this potential with respect to $\mathbf{v}$. The fixed points for the velocity are solutions of the equation $v_y(1 - v_y^2) - f = 0$. Figure~\ref{Fig2}(a) represents the value of these steady solutions, $v^\star$, as a function of the external force, $f$. Two regions can be identified and are separated by a critical value of the external force $f^\star = 2/(3\sqrt{3})$. For $f<f^\star$, two solutions are stable with respect to $v_y$, one parallel to the force field $v^-$ (dashed line) and one anti-parallel to the force field $v^+$ (solid line). The anti-parallel solution $v^+$ is unstable with respect to $v_x$, leading to a saddle node in the $(v_x,v_y)$ plane. The solution near the origin $(v_x,v_y)=(0,0)$ is unstable along both directions (dotted black line). For $f\gg f^\star$ the equation admits only one solution $v^-$ which is parallel to the force field ($v^-<0$) (solid line). \\
	
	We numerically solve Eqs.~\ref{Equation2D} for two asymptotic values of force fields, $f=10^{-3}\ll f^\star$ and $f=10^3 \gg f^\star$, and present in Fig.~\ref{Fig2}(b) the penetration depth, $\ell_c$, reached by the particle as a function of the incidence angle $\theta_0$. We observe a qualitative change in the behaviour close to $\theta_0 = 0$. The case of a classical Hamiltonian particle (gray line) has been superimposed. Indeed, Hamiltonian particles cannot travel beyond a critical penetration depth, $\ell_c^H =1/2$, indicated in dashed red line. As observed in Fig.~\ref{Fig2}(b), thanks to the Rayleigh friction, if $f<f^\star$ a barrier of potential energy larger than the kinetic energy $K_0 = m V_0^2/2$ can be overcome for small values of the incident angle $\theta_0$. For a harmonic force, a similar transition is observed in the inset of Fig.~\ref{Fig2}(b). The two regimes are separated by a critical angular frequency, $\omega^\star = 1/2$.\\
		
	The qualitative change of propulsion can be revisited by analyzing how the flow structure changes with $f$ in the velocity phase space [see Fig.~\ref{Fig2}(c) and (d)]. For $f = 0.10$ [see Fig.~\ref{Fig2}(c)], the flow is directed toward the unit circle and converges to the fixed point $v_y = v^-$; in the immediate vicinity of the unit circle. For $f = 0.70$ [see Fig.~\ref{Fig2}(d)], the flows converges directly toward $v_y=v^-$. This change of phase space topology derives from the collapse of the saddle point and the unstable fixed point $v^+$ at the critical value $f^\star$, as shown in Fig.~\ref{Fig2}(a). For $f < f^\star$, the velocity mainly changes in terms of orientation rather than in amplitude, while the opposite situation occurs for $f>f^\star$.\\
	
	 This transition can also be discussed by considering the different time scales governing the dynamics. As suggested by Eq.~\ref{equationofmotion}, the convergence to the unit circle $\Vert \mathbf{v}\Vert = 1$ originates from the self-propulsion and occurs over a time scale $\sim T_v$. The convergence to $v_y=v^-$ due to the force field occurs over a time scale $\sim T_F = (mv_0)/\mathcal{F}$. The ratio between the two time scales is given by $T_v/T_F = f$. Therefore, at low values of $f$, the system first converges to $\Vert \mathbf{v}\Vert = 1$ over a time $T_v$ and it aligns its velocity with the force field over a longer period of time ($T_F \gg T_V$). Therefore, for $f \ll f^\star$, signification penetration depths into the force field are possible, since the time spent with a velocity unaligned to $\mathcal{F}$ can be much larger than the typical time of interaction with the force field. As expected, motions in regions of high potential energy are therefore possible owing to the self-propulsive mechanism.\\
	
	\begin{figure}[t!]
		\begin{centering}
			\includegraphics[width=0.51\columnwidth]{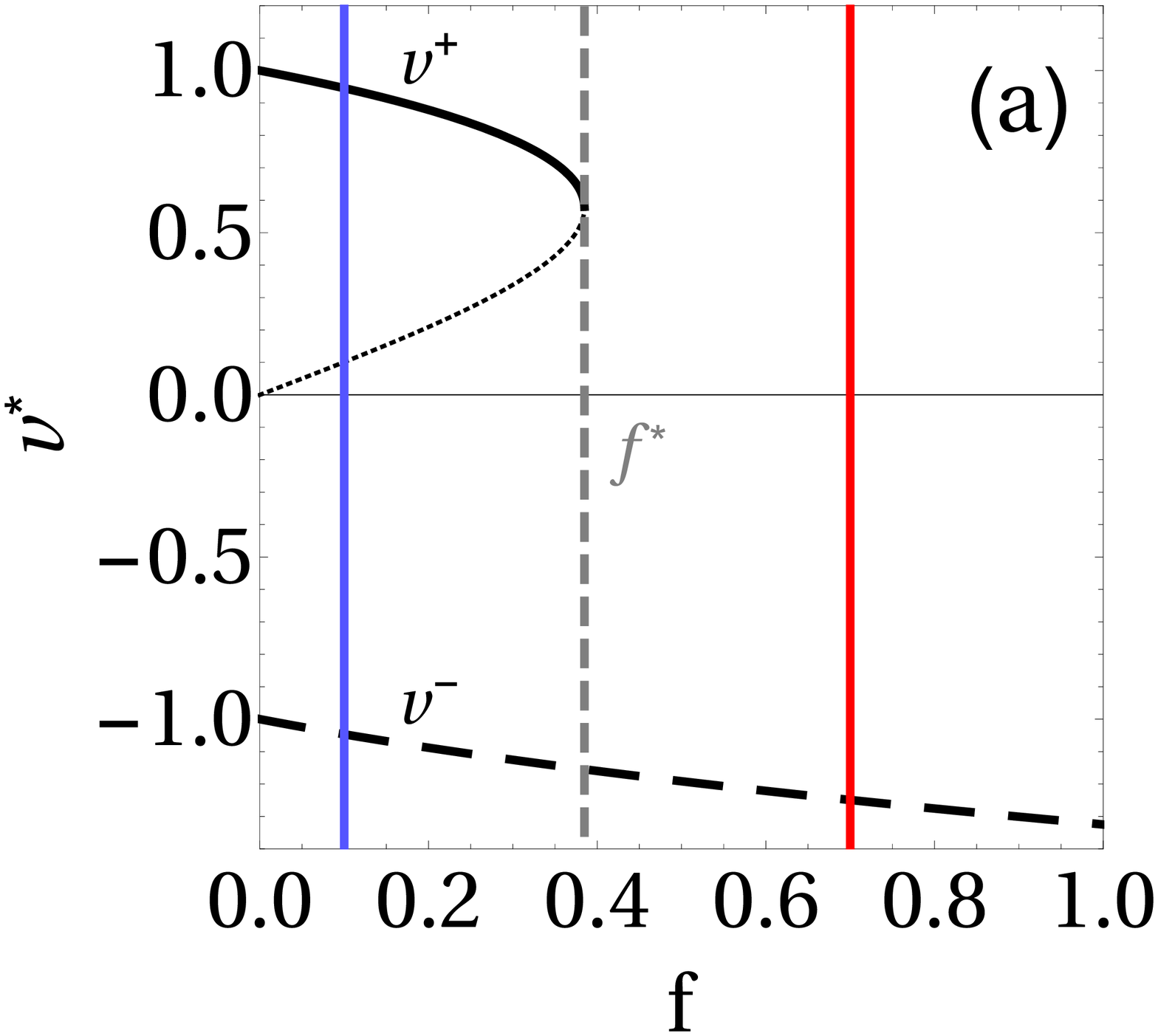}
			\includegraphics[width=0.47\columnwidth]{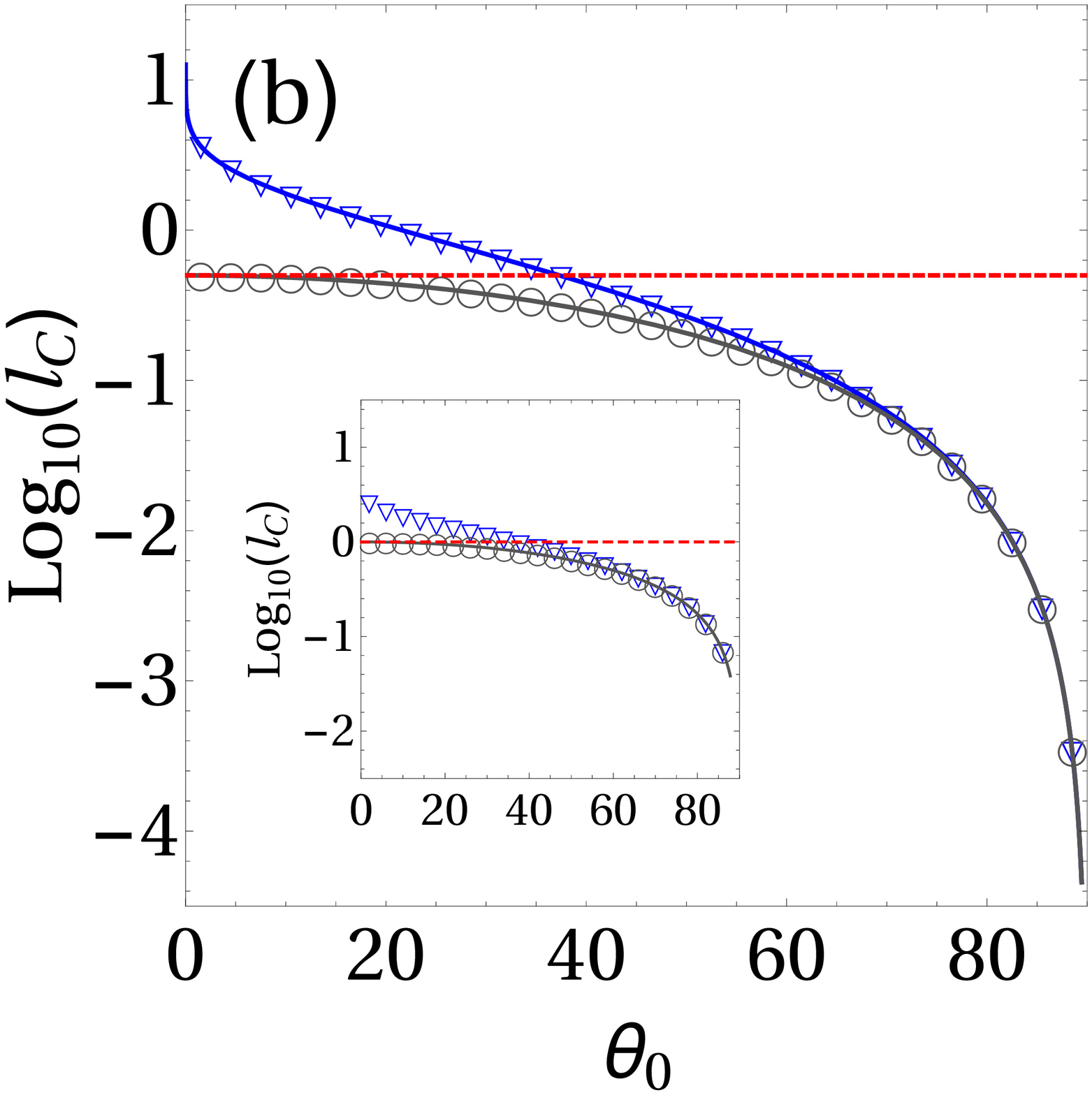}\\
			\vspace{5mm}
			\includegraphics[width=\columnwidth]{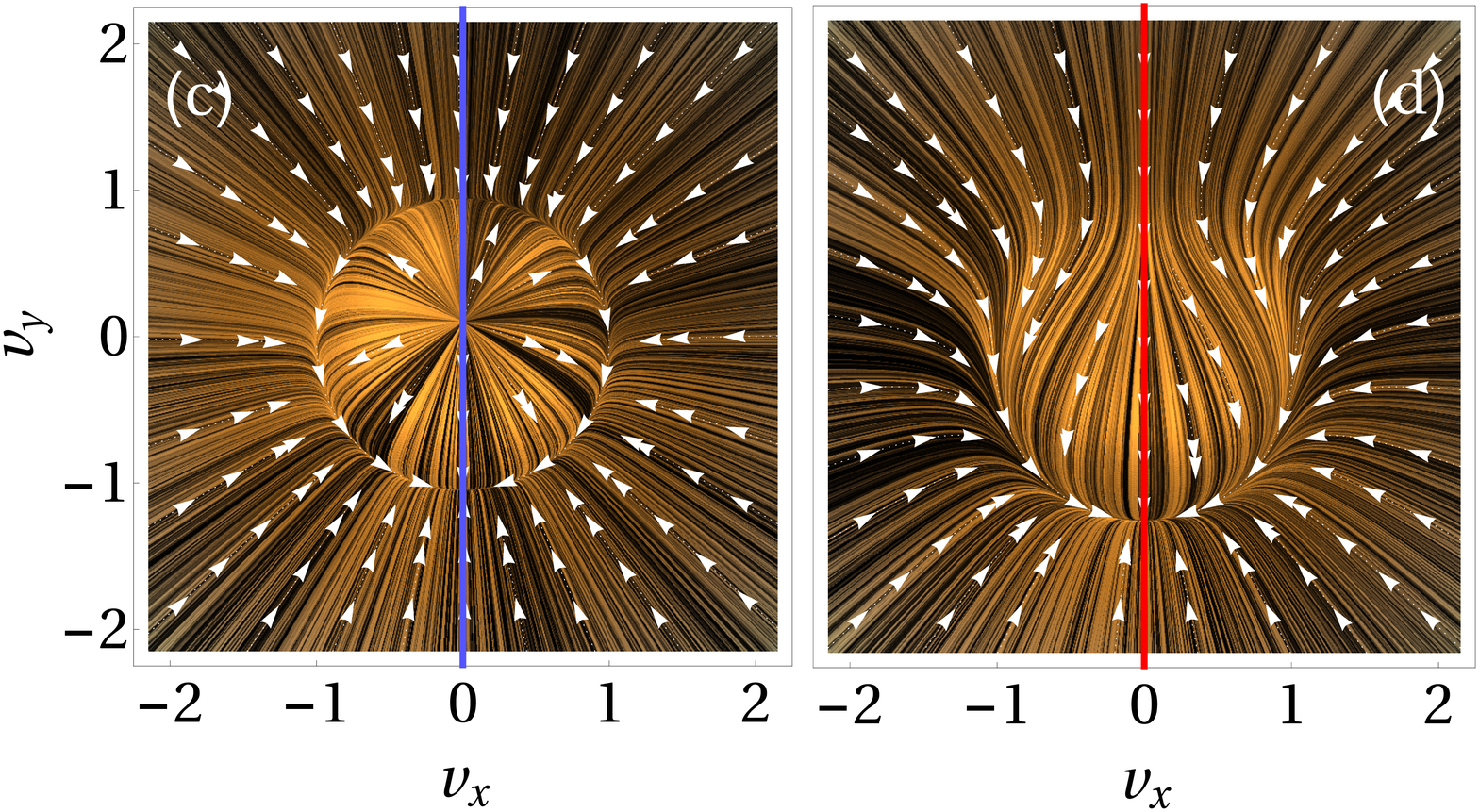}			
			\caption{\emph{Color online} (a) Equilibrium speed $v^\star$ of Eq.~\ref{Equation2D} for $v_x=0$ as a function of $f$. Stable solution $v^-$ (dashed black line), saddle node solution $v^+$ (solid black line) and unstable solutions (dotted black line). The critical force value $f^\star$ is defined graphically as the merging of the saddle-node and the unstable fixed point. (b) Penetration depth reached by the particle as a function of the angle of incidence $\theta_0$ (log scale). Low force regime $f=10^{-3}<f^{\star}$ ({\color{blue} $\triangledown$}) and high force regime $f=10^{3}>f^{\star}$ ({\color{mygray} $\circ$}). Solid gray line: maximal penetration depths for an Hamiltonian particle. Solid blue line: maximal penetration depth for a self-propelled particle in the limit $f\ll f^\star$ predicted by Eq.~\ref{thetaC}. Dashed red: limit encountered by Hamiltonian particles. Inset: Same analysis performed with a harmonic force field (identical color code). (c) and (d) flow representation in the ($v_x, v_y$)-plane respectively for $f=0.1<f^{\star}$ (blue line in Fig.~\ref{Fig2}(a)) and for $f=0.7>f^{\star}$ (red line in Fig.~\ref{Fig2}(a)) of Fig.~\ref{Fig2}(c) and (d). White arrows indicate the direction of the flow. }
			\label{Fig2}
		\end{centering}
	\end{figure}	
	
	We end this section by deriving an analytic expression for the penetration depth in the low force regime $f \ll 1$. In this regime, the penetration depth can be estimated by taking advantage of the separation of time scales between the fast dynamics of convergence to the unit circle $\vert \vert V \vert \vert=1$ and the slow dynamics of velocity direction change. Using this hypothesis, we look for a relation between the incidence angle, $\theta_c$, and the maximum dimensionless depth reached, $\ell_c$, as a function of the external force, $f$. The maximum penetration depth, $L_c$, reached in the force field can be expressed as 
\begin{equation}
	L_c = \int_0^{t_f} V_y dt,
\end{equation}
where $t_f$ denotes the time at which $V_x=0$. Due to the axial symmetry of the velocity phase space in the case $f\ll f^\star$, it is convenient to write eqs.~\ref{Equation2D} in cylindrical coordinates $(\Vert v\Vert, v_{\theta})$
\begin{equation}
\left\{
    \begin{array}{l} 
    \dot{\vert \vert \mathbf{v}\vert\vert}=\vert \vert \mathbf{v}\vert\vert\left(1-\vert \vert \mathbf{v}\vert\vert^2\right) - f\cos\theta,\\
    \vert \vert \mathbf{v}\vert\vert v_\theta=-f\sin\theta.
     \end{array}
\right.
\label{Equation2Dcylindric}
\end{equation}
In the case $f \ll f^\star$, we can approximate the velocity by $\vert \vert \mathbf{v}\vert\vert \simeq 1$. The maximal dimensionless depth can be written
\begin{equation}
	\ell_c = \int_{\theta_0}^0\cot\theta \,d\theta.
\end{equation}
It yields
\begin{equation}
\sin\theta_0 = \exp\left(-\ell_c\right)=\exp\left(-\frac{\mathcal{F}L_c}{mV_0^2}\right),\label{thetaC}
\end{equation}	
which links the maximal depth, $L_c$, reached in the external potential and the incidence angle, $\theta_0$. Equation~\ref{thetaC} reproduces well the numerical results in~\ref{Fig2}(b) (blue solid line). Note also that Eq.~\ref{thetaC} predicts the critical angle, $\theta_c$, defining the transition between crossing and reflected trajectories and corresponding to $L_c = L$.\\

In this section we have investigated the solutions of Eq.~\ref{Equation2D} and showed that there exists two distinct regimes. Particularly in the low force regime, i.e. $f \ll f^\star$, the particle can reach significant penetration depths in either a constant or an harmonic force field. The qualitative change of behaviour is conveniently traced out by looking at the velocity potential. We derive an expression for the penetration depth as a function of the incident angle. This penetration depth can be arbitrary large for small incident angle. Large penetrations are possible only if the particle has an initial velocity included in a cone of aperture $2\theta_c$. The model proposed in this section is deterministic and the initial conditions are fixed. We now analyze the influence of random initial conditions on the crossing properties.
	
\section{Crossing probability \label{SecIV}}	

	In this section, we introduce random incidence angles, which leads to a probability $\mathcal{P}$ to cross the potential barrier of dimensionless depth $\ell$. This stochastic ingredient aims at reproducing a random distribution of initial conditions with a maximum probability for normal incidence, as observed in the experiments of Eddi {\it et al}~\cite{Eddi2009}. Note that we do not consider the stochastic counterpart of Eq.~\ref{Equation2D}, but do investigate the deterministic Eq.~\ref{Equation2D} under random initial incidence angles. A uniform distribution of angles, as well as a centered gaussian distribution, have been investigated.  Considering either the uniform distribution, $P(\theta_0) = \mathcal{U}(\pi/2,\pi/2)$, or the gaussian distribution, $P(\theta_0) = \exp(-\theta_0^2/2\sigma^2)/\sqrt{2\pi\sigma^2}$, the inset of Fig.~\ref{Fig3} shows the probability to pass through a barrier of dimensionless maximal potential energy $u_{\mathrm{max}}$. The linear and the quadratic potential lead to $u_{\mathrm{max}} = \ell $ and $u_{\mathrm{max}} = \ell^2/2$, respectively. The probability, $\mathcal{P}$, is found to decrease exponentially with $u_{\mathrm{max}}$, leading to $\mathcal{P}(\ell) = \alpha\exp(-\beta u_{\mathrm{max}})$, as observed in the inset of Fig.~\ref{Fig3}. Figure~\ref{Fig3} shows the evolution of $\beta$ with the force field parameters ($f$ and $\omega$ respectively). Near the transition $f\lesssim f^\star$, $\beta$ depends on the force field parameters. As long as $f\ll f^\star$ or $\omega \ll \omega^\star$ respectively, $\beta = 1$ is constant. In this latter regime, reintroducing dimensions and evaluating the probability distribution yields
		
	\begin{figure}[t!]
		\begin{centering}
			\includegraphics[width= 0.9\columnwidth]{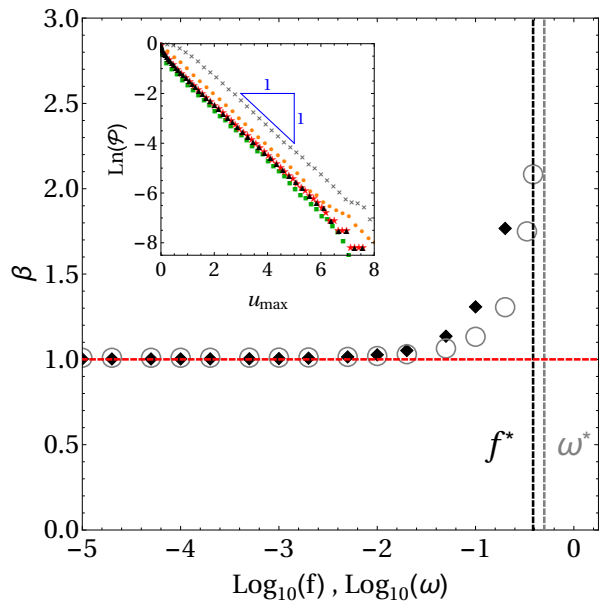}
			\caption{\emph{Color online} Evolution of the equivalent Boltzmann factor $\beta$ (log. scale along $x$) for a harmonic force field of natural frequency $\omega$  ({\color{mygray}\Large $\circ$}) and a constant force field of magnitude $f$ ({\footnotesize $\blacklozenge $}). Vertical dashed lines indicate $f^\star$ and $\omega^\star$ in gray and black respectively. {\bf Inset} Probability $\mathcal{P}$ to cross the potential barrier (log. scale) as a function of the maximal potential energy $u_{\mathrm{max}}$ of the energy potential barrier. Uniform distribution and harmonic force field ($\blacktriangle$), uniform distribution and constant force field ({\color{Red} $\bigstar$}). Gaussian distribution with $\sigma = \pi/8$ ({\color{mygray} $\times$}), $\pi/4$ ({\color{Orange} $\bullet$}), $\pi/2$ ({\color{ForestGreen} $\blacksquare$}) respectively and constant force field.}
\label{Fig3}
\end{centering}
\end{figure}

\begin{equation}
	\mathcal{P}(L) =  \alpha\exp{\left(-\frac{U_{\mathrm{max}}}{mV_0^2}\right)}.\label{Eq:Exponential}
\end{equation}
The expression of the probability distribution, $\mathcal{P}$, depends on the variance of the distribution, $P$, through the normalizing prefactor, $\alpha$, but not on the specific shape of the initial statistics $P(\theta_0)$ as shown in the inset of Fig.~\ref{Fig3} for various potentials. This scaling therefore appears as an general property of a Rayleigh-friction type of dynamics.\\

One may draw an analogy between our system and the canonical ensembles in statistical physics, in which the probability to cross a barrier of energy, $U$, is $\mathcal{P}= A \exp{\left(-U/k_BT\right)}$ and which leads to the following formal correspondence
\begin{equation}
 ``k_B T" = m V_0^2= 2 K_0.
\end{equation}
where $K_0$ is the passive kinetic energy. Knowing that the velocity of the particle is constrained by the Rayleigh friction for small values of $f$, this leads to an energy of $``k_BT/2"$ for the sole available degree of freedom, the direction of the instantaneous velocity.\\

In this section, we have shown that the model presented in Sec.~\ref{SecII} with random initial conditions in the limit $f \ll f^\star$ leads to crossing probabilities corresponding to a Boltzmann exponential law, but for a reason intrinsically different from a stochastic process. This result arises from the qualitative change in the phase space and the constraint $v=1$ for small external potentials. 
\begin{figure}[t!]
			\includegraphics[width=\columnwidth]{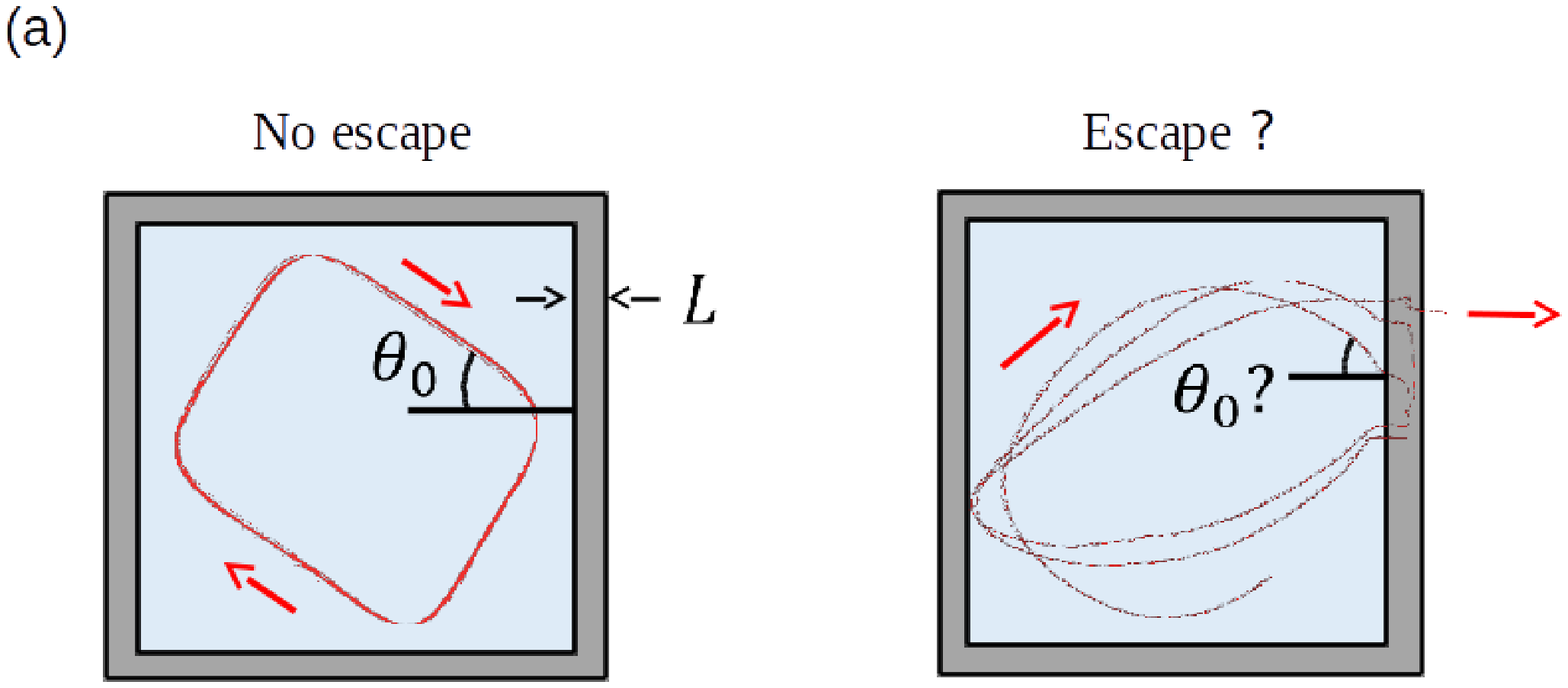}\\
			\includegraphics[width=0.90\columnwidth]{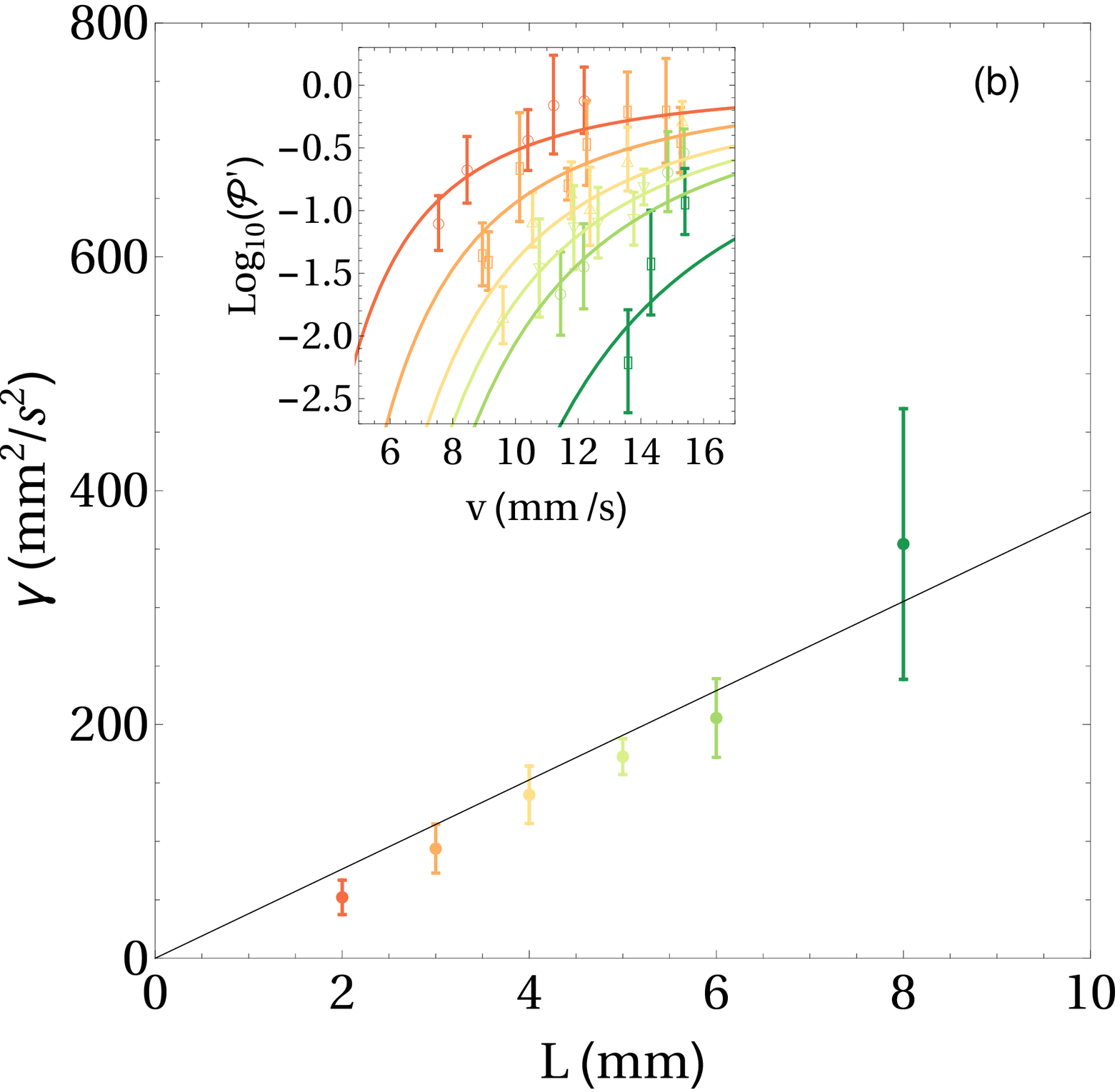}
			\caption{\emph{Color online} (a) Sketches of the tunneling experiment adapted with courtesy from Eddi et al. [Phys. Rev. Lett. 102, 240401 (2009)]. (Left) A bouncing drop is travelling along a limit cycle in a rectangular cavity. (Right) The impact angles are erratic and the walker crosses the barrier of potential after several reflexions. (b) Evolution of the potential shape $\gamma$ as a function of the obstacle thickness $L$ computed from the probability of crossing and the Eq.~\ref{Eq:fitExp}. Different colors indicate different thickness. \textbf{Inset} Probability to cross a barrier of given thickness as a function of the incoming droplet velocity. Color indicates the thickness, relative to the color code used in the main plot.}\label{Fig4}
	\end{figure}
	
\section{Comparison with the experimental data \label{SecV}}	
	
	 This model can be applied to experiments in which self-propelled particles are confined and interact with slow variating potentials. Such a situation has been encountered by Eddi \textit{et al.}~\cite{Eddi2009} with self-propelled droplets bouncing on an air-water interface. In this section we compare our theoretical predictions with the existing data from Eddi \textit{et al.}~\cite{Eddi2009}.\\
	 
	 A sketch of the experiment is drawn in Fig.~\ref{Fig4}a. Repeated drop impacts on the fluid surface create a standing Faraday wave field pattern~\cite{Eddi_JFM_2011,Molacek2013}, which in return propels the drop~\cite{Couder2005,Bush2015,Filoux2015}. This dual system is termed a \emph{walker}. Eddi \textit{et al.}~\cite{Eddi2009} performed the following experiment. A single drop was trapped in rectangular or rhomboidal cavities separated by submarine walls. These walls repelled walkers and thus acted as barriers of potential. \\
	 
	 The precise description of the drop/wall interaction was not known for several years. Recently, it was shown that the shape of the effective potential can be modelled~\cite{Faria2017,Nachbin2016,Pucci2016}. Subsequently one-dimensional crossings were investigated~\cite{Nachbin2016} and rationalized, but the two-dimensional situation remained a numerical and theoretical challenge. As our results (Eq.~\ref{Eq:Exponential}) do not depend on the exact shape of the repulsive potential, it is tantalizing to apply our model to this situation. In the experiments, the erratic crossing events originate from the interaction with the propelling waves and memory effects, which are known to trigger a transition to chaos as soon as the drop interacts with external potential~\cite{Perrard_PRL_2014,Tambasco2016}. The chaotic regime generates an effective distribution of incident angles. Under these assumptions, according to Eq.~\ref{Eq:Exponential}, the experimental probability, $\mathcal{P}$, to cross the barrier should read
\begin{equation}
	\mathcal{P}(L,V_0) = \exp\left(-\frac{\gamma(L)}{V_0^2}\right), 
	\label{Eq:fitExp}
\end{equation}
where $\gamma(L)$ is a potential that depends on the thickness of the submerged barrier, $V_0$ is the free walker speed and $\gamma(L)$ is the sole fitting parameter. In Fig.~\ref{Fig4}(b) we compare our predictions with the experimental data of Eddi {\it et al.}~\cite{Eddi2009}. Specifically, we show the evolution of $\gamma$ as a function of the thickness $L$ of the barrier. We observe that $\gamma$ changes linearly with $L$ in accordance with the existing experimental data (coefficient of determination $R^2 = 0.981$). This linear dependency indicates that the submarine walls confining the walker are well-described by an effective step force field. In the article of Eddi {\it et al.}~\cite{Eddi2009}, this result was attributed to the lowest excitability of the Faraday waves above the submarine obstacle. Finally the inset of Fig.~\ref{Fig4}b shows the experimental and theoretical probabilities to cross a barrier of given thickness as a function of the incoming velocity of the walker. As observed from the inset of Fig.~\ref{Fig4}b, Eq.~\ref{Eq:fitExp} correctly fits the experimental data. \\

So far only the erratic distribution of impact angles has been considered, which arises from the wavelike properties of the system. We also show that the self-propulsion mechanism itself is sufficient to rationalize the crossing properties observed by Eddi {\it et al.}~\cite{Eddi2009}. Note also that the current model for self-propulsion best holds for short wave damping time, a regime in which these drops can simply be seen as self-propelled particles~\cite{Labousse2014} with an added effective mass~\cite{Bush2014}. But even for more complex regimes, the non-Hamiltonian structure of the dynamics will impose strong constraint on the walker tangential force balance and thus be of some general relevance.  \\
	
\section{Conclusions \label{SecVI}}

	Non-Hamiltonian particles can travel through energy barriers thanks to their self-propulsion mechanism. This property is strongly connected to the velocity phase space topology. The details of the flow structure depend on the type of external force, but the transition between low and high force regimes will remain for various force fields. In this article, we leverage this non-Hamiltonian property to rationalize the experiments of walker tunneling carried out by Eddi {\it et al}~\cite{Eddi2009}. In accordance with previous theoretical investigations~\cite{Labousse2014,Bush2014}, we propose a theoretical model in Sec.~\ref{SecII}. Then in Sec.~\ref{SecIII} we perform a stability analysis of the fixed points and show the existence of these two regimes. In Sec.~\ref{SecIV}, a lack of information about the initial conditions is incorporated (here, the initial angle of incidence) and shown to lead to a probabilistic point of view. In the low force regime, this creates a dynamics reminiscent of thermally activated systems. Finally in Sec.~\ref{SecV}, we apply our model to the experimental case of self-propelled drops. Our investigation is a crucial step in understanding the tunneling of walkers. Indeed we show that the non-Hamiltonian self-propulsion properties of walkers are sufficient to rationalize their crossing of submarine barriers of potential. However, we introduce the randomness of initial conditions as an {\it ad hoc} ingredient. Understanding the origin of this complexity requires one to account for the wave-like nature of the walkers' propulsion. This second step is beyond the scope of this article, but the recent effective depth models~\cite{Faria2017,Pucci2016} are promising methods for investigating the onset of scattered distribution of initial angles. \\

\section*{Acknowledgments}
This work was financially supported by the Actions de Recherches Concert\'ees (ARC) of the Belgium Wallonia-Brussels Federation under Contract No. 12-17/02. M.L. and S.P. acknowledge the financial support of the French Agence Nationale de la Recherche, through the project ‘ANR Freeflow’, LABEX WIFI (Laboratory of Excellence ANR-10-LABX-24), within the French Program ‘Investments for the Future’ under reference ANR-10-IDEX-0001-02 PSL. The authors thank warmly Antonin Eddi for sharing experimental data and precious advices, thank Vincent Bacot and Emmanuel Fort for insightful discussions and Nicolas Vandewalle and Yves Couder for fruitful discussions and careful readings: M.L. thanks Scott Waitukaitis for his careful reading.


\begin{thebibliography}{10}

\bibitem{Marchetti2013} Marchetti M. C., Joanny J.~F., Ramaswamy S., Liverpool T.~B., Prost J., Rao M., and Aditi Simha R. \emph{Rev. Mod. Phys.} \textbf{85}, 1143 (2013).

\bibitem{Solon2015} Solon A.~P., Caussin J.~B., Bartolo D., Chat\'{e} H. \& Tailleur J. \emph{Phys. Rev. E} {\bf 92}, 062111 (2015).

\bibitem{Deseigne2010} Deseigne J., Dauchot O. \& Chat\'{e} H. \emph{Phys. Rev. Lett.} {\bf 105}, 098001 (2010).

\bibitem{Bricard2013} Bricard A., Caussin J.~B., Desreumaux N., Dauchot O. \& Bartolo D. \emph{Nature (London)} {\bf 503}, 95 (2013).

\bibitem{Rayleigh1877} Rayleigh J.~W.S. \emph{The theory of sound} (Macmillan and Co 1877).

\bibitem{Erdmann2005} Erdmann U. \& Ebeling W. \emph{Int. J. Bifurcation Chaos} {\bf 15}, 3623 (2005).

\bibitem{Bechinger2016} Bechinger C., Di Leonardo R. L\"{o}wen H., Reichhardt C., Volpe G. \& Volpe G. \emph{Rev. Mod. Phys.} {\bf 88}, 045006 (2016).

\bibitem{Romanczuk_EPJ_2012} Romanczuk P., B\"{a}r M., Ebeling W., Lindner B. \& Schimansky-Geier L. {\emph Eur. Phys. J. Special Topics} {\bf 202}, 1-162 (2012).

\bibitem{Erdmann_2003} Erdmann U. \& Ebeling W. {\emph FNL} {\bf 3}, 145 (2003).

\bibitem{Kearns_2010} Kearns D. B. {\emph Nat. Rev. Micro.} {\bf 8}, 634 (2010).

\bibitem{Lindner2008} Lindner B. \& Nicola E.M. \emph{Eur. Phys. J. Special Topics} \textbf{157}, 43 (2008).

\bibitem{Erdmann2000} Erdmann U., Ebeling W., Schimansky-Geier L. \& Schweitzer F., \emph{Eur. Phys. J. B.} \textbf{15}, 105 (2000).

\bibitem{Burada2012} Burada P.S. \& Lindner B. \emph{Phys. Rev. E} \textbf{85}, 032102 (2012).

\bibitem{Schweitzer2000} Schweitzer F., Tilch B. \& Ebeling W. \emph{Eur. Phys. J. B} {\bf 14}, 157 (2000).

\bibitem{Eddi2009} Eddi A., Fort E., Moisy F. \& Couder Y. \emph{Phys. Rev. Lett.} {\bf 102}, 240401 (2009).

\bibitem{Nachbin2016} Nachbin A., Milewski P. \& Bush J.~W.M., \emph{Phys. Rev. Fluids}{\bf 2} (3), 034801 (2017)

\bibitem{Faria2017} Faria L.M.J. \emph{ J. Fluid Mech.} {\bf 811}, 51 (2017).

\bibitem{Pucci2016} Pucci G., S\'{a}enz P.~J., Faria L.~M. \& Bush J.~W.M., \emph{ J. Fluid Mech.}{\bf 804} R3 (2016).

\bibitem{Labousse2014} Labousse M. \& Perrard S. \emph{Phys. Rev. E} \textbf{90}, 022913 (2014).

\bibitem{Bush2014} Bush J.~W.M., Oza A.~U. \& Mol\'{a}\v{c}ek J. {\emph J. Fluid Mech.} {\bf 755}, R7 (2014).

\bibitem{Helbing2001} Helbing D. \emph{Rev. Mod. Phys.} {\bf 73}, 1067 (2001).

\bibitem{Eddi_JFM_2011} Eddi A., Sultan E., Moukhtar J., Fort E. Rossi M. \& Couder Y. {\emph J. Fluid Mech.} {\bf 674} 433-463 (2011).

\bibitem{Molacek2013} Mol\'{a}\v{c}ek J. \& Bush J.~W.M \emph{J. Fluid Mech.} {\bf 727} 612 (2013). 
  
\bibitem{Couder2005} Couder Y., Proti\`{e}re S., Fort E. \& Boudaoud A. \emph{Nature (London)} {\bf 437}, 208 (2005).

\bibitem{Bush2015} Bush J.~W.M, \emph{Ann. Rev. Fluid Mech.} {\bf 49}, 269 (2015).

\bibitem{Filoux2015} Filoux B., Hubert M. and Vandewalle N. \emph{Phys. Rev. E}  {\bf 92}, 041004(R) (2015). 

\bibitem{Perrard_PRL_2014} Perrard S., Labousse M., Fort E. \& Couder Y. \emph{ Phys. Rev. Lett.} {\bf 113} 104101 (2014).

\bibitem{Tambasco2016} Tambasco L.D., Harris D.M., Oza A.U., Rosales R.R. \& Bush J.W.M. \emph{Chaos} \textbf{26} 103107 (2016).

\end{thebibliography}
\end{document}